\documentclass[journal,onecolumn,12pt]{IEEEtran}

\usepackage{multirow}
\usepackage{graphicx}
\usepackage{color}
\usepackage{subfigure}
\usepackage{epsfig}
\usepackage{citesort}
\usepackage{setspace}
\usepackage{latexsym}
\usepackage{amssymb}
\usepackage{amsmath}

\title{Performance of a random-access wireless network with a mix of full- and half-duplex stations}
\author{Vaneet Aggarwal and N.~K.~Shankaranarayanan \thanks{ The authors are with AT\&T Labs-Research, Florham Park, NJ 07932, email: \{vaneet,shankar\}@research.att.com }}
\newtheorem{theorem}{Theorem}

\begin{document}
\maketitle

\begin{abstract}

In this paper, we consider the performance of a random-access time-slotted wireless network with a single access point and a mix of half- and full- duplex stations. Full-duplex transmissions involve data transmitted simultaneously in both directions, and this influences the dynamics of the queue at the access point. Given the probabilities of channel access by the nodes, this paper provides generalized analytical formulations for the throughputs for each station. Special cases related to a 802.11 DCA based system as well as a full-fairness system are discussed, which provide insights into the changes introduced by the new technology of full-duplex wireless.

\end{abstract}

\section{Introduction}

Currently deployed wireless communications systems cannot transmit and receive on the same frequency at the same time, i.e.,\ networks do not operate in a single-channel full-duplex fashion. As a result, the networks are either time-division duplex (e.g.,\ WiFi) or frequency-division duplex (e.g.,\ UMTS cellular). The key challenge in achieving true full-duplex communication is the large power differential between the ``self-interference" created by a node's own radio transmission and the signal of interest. Recently, there has been a lot of progress in new full-duplex designs \cite{Melissa1,MelissaTVT,sachin2,sachin1,khandani,msft,chiang,hotsnets} at the physical layer design of full-duplex systems. In this paper, we consider a wireless network in an infrastructure mode with one access point and a mix of full- and half- duplex capable stations, where there is traffic flowing only between the access point and the stations, and there is no peer-to-peer traffic between stations. This paper characterizes the throughput to/from each station.

Besides the single access point (AP), let there be $m$ full-duplex and $n$ half-duplex stations in the system. We consider a time-slotted system in which each node grabs the channel with a certain probability as in a distributed random access system. Full-duplex transmission in a random-access system requires additional signalling. The MAC protocol in \cite{MelissaTVT} uses RTS/CTS signalling, with the advantage that the collisions only happen for RTS packets which are much smaller than data packets. Based on this observation, we ignore the effect of collisions in our throughput calculations. We allow for the AP and the two types of nodes to access the channel with different probabilities. For the special case of 802.11 distributed channel access (DCA), the backoff timing process would determine these probabilities \cite{updownfair}. We consider a generalized system where there may be priorities. Thus, each of the three categories of node (access point, full-duplex station, half-duplex station) can have different access probabilities. We assume that all communications are between the access point and stations, i.e. there are no peer-to-peer communications. We also assume saturated traffic where the buffers are always full. During a full-duplex communication, the access point simultaneously receives from, and transmits to, a full-duplex station. During a half-duplex communication, the access point either receives from, or transmits to, a full-duplex station. A MAC design based on this assumption which demonstrates that the full- and half- duplex nodes can  coexist in a 802.11 based system was provided in \cite{MelissaTVT}. The introduction of full-duplex wireless technology changes the protocol dynamics in a fundamental way, and this paper generalizes the problem to consider general channel access probabilities for the three types of nodes.

We assume that the access point has a single queue for the data packets destined for the stations, and that the destination of each arriving packet is equally likely to be any of the stations. When a full-duplex station grabs (wins the contention process to initiate transmission) the channel to send data to the AP, the first packet in the AP queue that is destined for this station is removed and a full-duplex transfer happens between the AP and the full-duplex station as in \cite{MelissaTVT}. This leads to full-duplex packets being removed from the access point queue out of turn. When the access point grabs the channel, it sends data to the station to which the head of the queue data is addressed. Since many out of turn packets for full-duplex have been removed, it is not readily obvious what proportion of the time slots grabbed by access point are used to send data to a half-duplex node vs. a full-duplex node. In this paper, we provide analytical formulations for the probability of sending data from the AP to any station when the AP grabs the channel.

We consider two special examples to illustrate the use of the equations. The first is a 802.11 DCA system, where all the nodes have an equal chance of channel access. This system is more suitable for uplink than downlink for half-duplex stations since the sum downlink to all half-duplex stations decreases as the number of half-duplex stations increases even though the sum uplink does not decrease. In teh steady state for such a system, we find that when the AP grabs the channel, it never transmits to any full-duplex station and thus these time-slots are used for downlink to half-duplex stations with probability 1. We further note that the downlink throughput to a half-duplex station is much less than the uplink throughput for the half-duplex station as per the inherent nature of 802.11 DCA system for symmetric uplink/downlink traffic. To improve the symmetry and fairness, we consider a second special case where the uplink and downlink throughput for each station (whether half- or full- duplex) is the same. If there were no full-duplex stations, this corresponds to half the time for uplink and half the time for downlink and this is equivalent to the AP grabbing the channel for half of the time slots. With the increase of full-duplex stations in the mix, the fraction of the slots used by the AP changes appropriately, and there is a gradual increase of the sum (normalized) throughput from 1 to 2 while maintaining fairness. We see that the fairness among the nodes and up/down directions comes at a moderate expense of the total system throughput for a system with a mix of full- and half- duplex nodes.

The rest of the paper is organized as follows. Section II describes the system model. Section III describes the main result. In Section IV, we consider two special cases of the result. Section V relates the results to the MAC implementation in \cite{MelissaTVT} and Section VI concludes the paper.


\section{System Model}
In this paper, we consider a single Access Point (AP) serving a mixture of full-duplex and half-duplex capable stations. The communication system is assumed to be time-slotted with distributed random access. The AP is assumed to have full-duplex capability, i.e., it can transmit and receive simultaneously. The half-duplex stations cannot transmit and receive at the same time. In a  time-slot, a half-duplex station can either transmit or receive, but not both. A full-duplex station can transmit and receive simultaneously in a time-slot like the AP.

Besides a single AP, we assume there are $m$ full-duplex stations and $n$ half-duplex stations. We assume that the AP has a single queue which is filled by packets destined with equal probability for each station, and each station has packets only for the AP. Further, we assume that all the buffers are full and thus all the nodes (AP and stations) always have data to send. With the random-access scheme, there is a contention in each time-slot to decide which node grabs the channel in that time slot to be able to initiate transmission. Depending on priority schemes, the channel access probabilities may be different for different nodes. Let the AP get the channel for $p_A$ fraction of the time slots. Further, we assume that a full-duplex station wins the channel access for $p_F$ fraction of the time slots and a half-duplex station wins the channel access for a $p_H$ fraction of the time slots. Since each time slot is used by some node, we have $p_A+mp_F+np_H=1$.

We assume the following system operation: There is a single queue at the AP which serves all the stations. Further, packets arriving at the AP have a destination which is equally likely to be any of the stations. The size of the queue is large enough that it never becomes a bottleneck. When the AP wins the contention for a time slot, it sends the packet which is at the head of its queue. If the packet at the head of the queue is for a half-duplex station, the time-slot has a half-duplex transmission of data from AP to the half-duplex station. If the packet at the head of the queue is for a full-duplex station, the time-slot is a full-duplex time-slot in which AP sends data to, and receives data simultaneously from, the particular full-duplex station. If a half-duplex station wins the contention in a time slot, there will be a half-duplex communication from the station to the AP. If a full-duplex station wins the contention in a time slot, we assume that the AP searches for the first packet in its queue for the full-duplex station and sends data (if any) to the station while simultaneously receiving data from the station. Thus, when a full-duplex station wins the contention, the AP may send a packet out of turn as part of the response to incoming data from a full-duplex station. We assume that each time slot involves random access, the AP and stations can segment packets as needed and send data in time slots, and that packets from the AP to a station use an integer number of time slots.

\section{Main Result}
In this section, we will derive a generalized analytical formulation for the system throughputs. We normalize the achieved throughput in terms of the number of time slots in which a node sends data. This can be mapped into bits per second based on the link efficiency, and the results can thus be applied to rate-adaptive systems. More precisely, a throughput from node A to node B of $t$
implies that node A sends data to node B for $tN$ time-slots out of a total of $N$ time-slots as $N$ goes to infinity. In other words, the normalized throughput from node A to node B is the fraction of the total time-slots during which data is transmitted from node A to node B. Consider as an example a system with one AP and one station. If the station is half-duplex, the sum throughput from the station to the AP and from the AP to the station is 1 (one packet per slot, if we assume that each packet fits in one slot). However, this sum is 2 if the station is full-duplex since there is a packet in each direction during a time slot (two packets per slot). These sum throughputs would also be maintained if the system has multiple stations, as long as they are all full-duplex stations or all half-duplex stations. As full-duplex technology is developed and introduced, we envision full-duplex capable access points serving a mix of full-duplex stations and legacy half-duplex stations. The sum throughput will be between and 1 and 2. Our result below gives the different throughputs between AP and stations for the general case.

\begin{theorem}
In a stationary system with one AP, $m$ full-duplex and $n$ half-duplex stations where the AP, a full-duplex stations, and a half-duplex station wins the channel access with probability $p_A$, $p_F$, and $p_H$ respectively, the throughputs between various nodes are as follows.
\begin{eqnarray}
\text{AP to one HD station} && \frac{p_A}{n} p, \nonumber \\
\text{AP to one FD station} && \frac{p_A}{m} (1-p) + p_F, \nonumber  \\
\text{one FD station to AP} && \frac{p_A}{m} (1-p) + p_F, \nonumber  \\
\text{one HD station to AP} && p_H,  \nonumber
\end{eqnarray}
where $p=\min(1, \frac{n}{n+m}\frac{p_A+mp_F}{p_A})$ is the fraction of the times that the AP transmits to a half-duplex station after grabbing the channel. The total sum throughput is $1+m p_F+p_A(1-p)$.
\end{theorem}

The rest of the section gives the proof of this Theorem.

We first consider two special cases of a complete full-duplex system and a complete half-duplex system. If all the stations are half-duplex, $m=0$, $p_F=0$ and thus $p=1$. Thus, the total throughput $1+m p_F+p_A(1-p) = 1$ which is expected. If all the stations are full-duplex, $n=0$, $p_H=0$ and thus $p=0$. Using the fact that the sum of probabilities $np_H+mp_F+p_A=1$, we can sum all the throughputs and derive the sum throughput to be $1+mp_F+p_A=2$ which is expected. Having verified the two corner cases, we now prove the theorem for a system with a mixture of full- and half- duplex stations.

We first consider the case when $p_F> p_A/n$. In this case, the data packets for a full-duplex station get removed from the AP queue much faster than $p_A/n$ which means that even if the data packets for a half-duplex station are depleted at rate of $p_A/n$ from the AP queue, there are no FD data packets at the head of the AP queue with probability 1. This is because FD packets get depleted at higher rate than HD packets thus leaving no FD packets at the head of the AP queue with probability 1. This means that when AP gets a chance to start a transmission it only sends data to HD stations. Given that $p$ is be the probability that AP sends data to any HD station when it grabs the channel, we have $p=1$.

If $p_F\le p_A/n$, we can see that the probability that AP sends data to any HD station when it grabs the channel given by $p \le 1$. Then, the throughput from AP to a HD station is $pp_A/n$ while the throughput from AP to a FD stations is $p_F+(1-p)p_A/m$. If  $pp_A/n>p_F+(1-p)p_A/m$, the HD packets will be depleted faster from the AP queue thus giving FD data packets at the head of the queue. This will decrease $p$. However, if $pp_A/n<p_F+(1-p)p_A/m$, there will be more HD packets left at the head increasing $p$. Thus, the system stabilizes such that $pp_A/n=p_F+(1-p)p_A/m$. This gives $p=\frac{n}{n+m}\frac{p_A+mp_F}{p_A}$.

Thus, combining the two cases and assuming $m,n>0$,
we have $p=\min(1, \frac{n}{n+m}\frac{p_A+mp_F}{p_A})$.
We also state that the throughput from HD station to AP is $p_H$. This is because a HD station sends to the AP only when the HD station wins the channel access. The AP sends to a HD station when AP wins the channel and sends to a HD station with probability $p/n$. Thus, the throughput from the AP to a HD station is  $\frac{p_A}{n} p$. The throughput from AP to a FD station is the same as the throughput  from a FD station to AP. AP sends data to a FD station in two scenarios. The first is when a FD station grabs the channel, which happens in $p_F$ fraction of the time-slots. The second is when AP grabs the channel and has data for this FD station at the head of the queue which happens in  $\frac{p_A}{m} (1-p)$ fraction of the time-slots. Thus, the throughput from AP to FD station is  $\frac{p_A}{m} (1-p) + p_F$ which proves the statement of the theorem.

\section{Examples and Numerical Results}
In this section, we consider two special cases of the main result. The first case is a 802.11 based system where all nodes get equal channel access while the second system adjusts the channel access probabilities to achieve fairness.

The first special case is a 802.11 DCA based system where we assume that each frame (packet) fits in one time slot, and each node has equal chance of winning a time-slot. Thus, $p_A=p_F=p_H$. We assume both $n$ and $m$ are greater than $0$. 
In this case, from the theorem, we have that $p=\min(1, \frac{n}{n+m}\frac{p_A+mp_F}{p_A}) = \min(1, \frac{n+nm}{n+m}) = 1 $. 
This means that when AP gets the channel, it sends data to only half-duplex stations. Thus, the throughput to a HD station is $p_A/n$, throughput from a HD station to the AP is $p_H$, and the throughput to a FD station is $p_F$ with $p_A=p_F=p_H=\frac{1}{1+m+n}$. The MAC design in \cite{MelissaTVT} for instance achieves $p_A=p_F=p_H$, and it was noted that in the steady state the AP has no data for a FD station when AP grabs the channel. The  sum throughput of the system in this case is $1+\frac{m}{1+m+n}$. Thus, the  sum throughput increases by a factor of $1+\frac{m}{1+m+n}$ as compared to a legacy half-duplex system. As an example, we consider 40 stations, where $m$ of them are full-duplex stations. Figure 1(a) depicts the total uplink (to the AP) and downlink (from the AP) throughputs to both full- and half- duplex stations and the total system throughput as a function of $m$. The total system throughput increases with increase of full-duplex stations. We note that the uplink and downlink throughputs for half-duplex stations are very asymmetric which is a characteristic of an infrastructure 802.11 based system (traffic only to/from AP) with saturated traffic.

\begin{figure*}
\centering
\mbox{\subfigure{\includegraphics[width=0.33\textwidth]{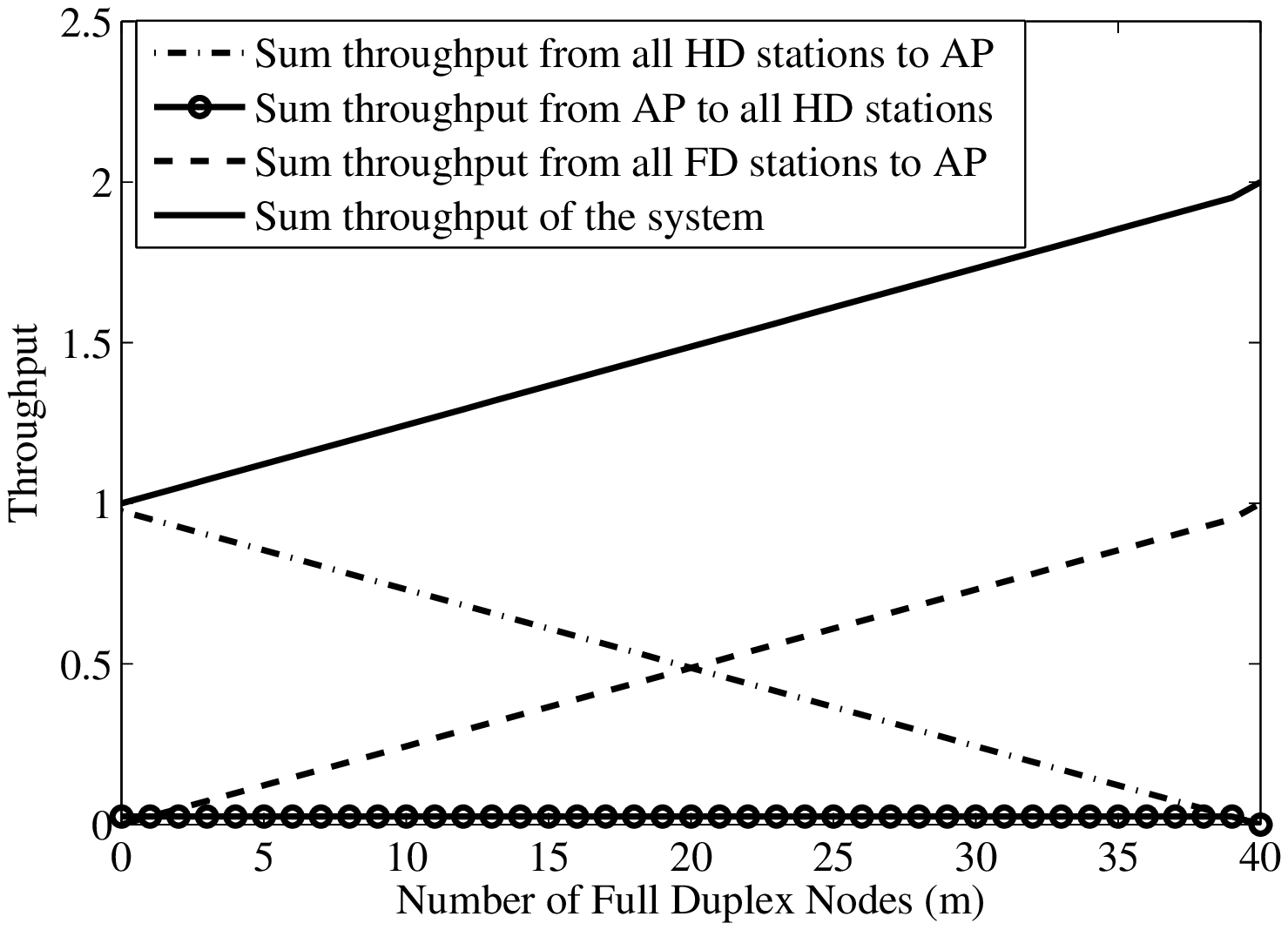}
\quad
\subfigure{  \includegraphics[width=0.32\textwidth]{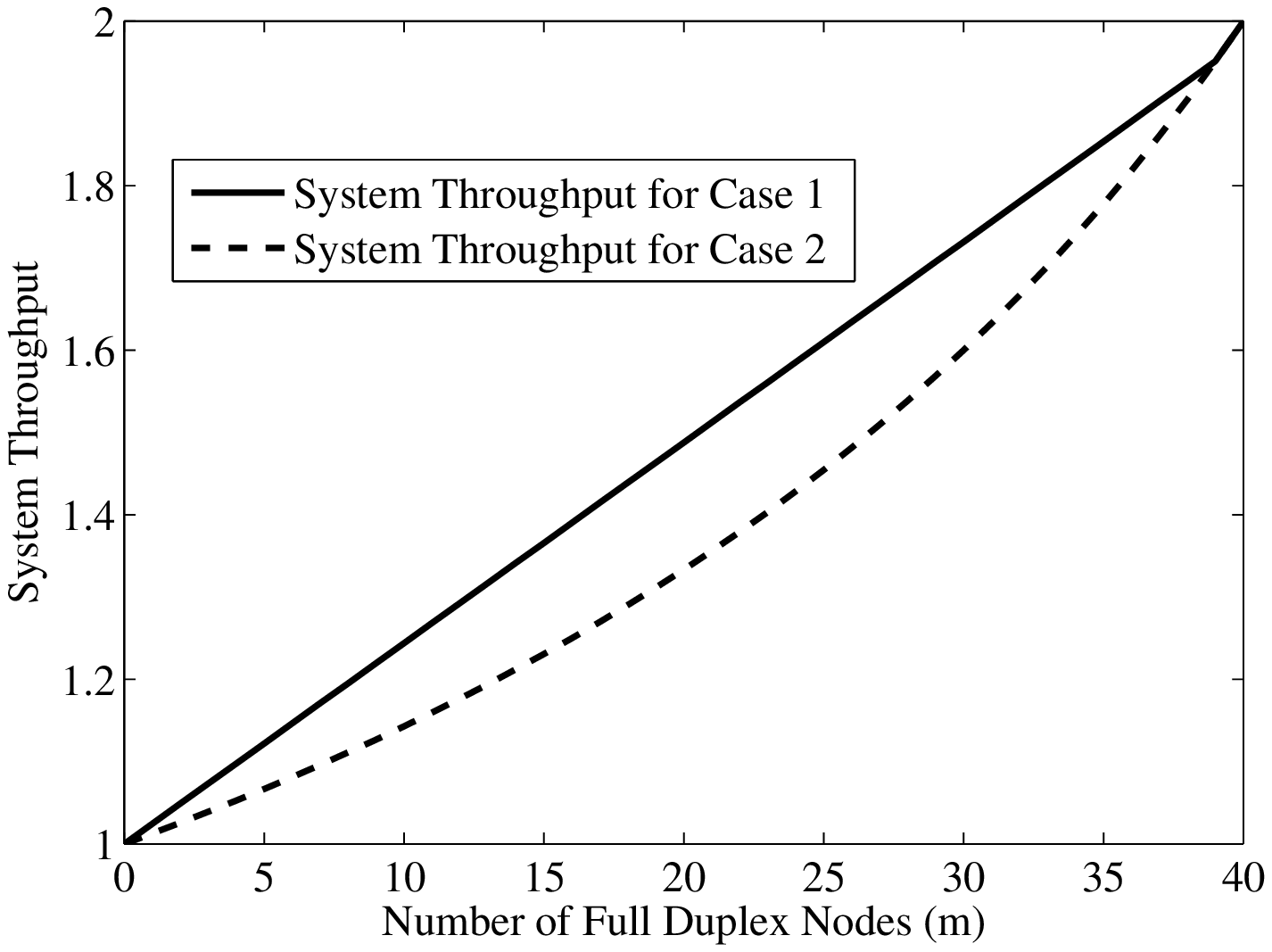} }}}
\caption{(a) Normalized throughput between nodes for the first special case (802.11 based DCA) for a system with $40$ stations, (b) Normalized System throughput for $40$ stations the two special cases: (1) 802.11 based DCA (2) system with full fairness for up and down flows for both full- and half-duplex stations}
\label{fig12}
\end{figure*}

%

\begin{table*}[ht]
\centering
\begin{tabular}{|c|c|c|c|c|}
\hline
 &    &  \multicolumn{3}{|c|}{Average Goodput (Mbps)}    \\
 \hline
Scenarios &   Sum Goodput   &   AP to HD node  &   AP to FD node,   &   HD node to AP \\
 &    (Mbps)            &                           & FD node to AP      &                        \\
 &     &   (HD downlink)  &   (FD downlink=uplink)  &   (HD uplink) \\

\hline
\hline
\multicolumn{5}{|c|}{$m=1$}\\
\hline
$2m$ FD  &  26.07 & -  & 6.52  &  -   \\
$2m$ HD   &  12.82  &  2.12 & -  & 4.29 \\
$m$ FD, $m$ HD & 17.87   & 3.81  & 4.81  & 4.45  \\
\hline
\hline
\multicolumn{5}{|c|}{$m=2$}\\
\hline
$2m$ FD  &  25.99 & -  & 3.25  &  -   \\
$2m$ HD   &  12.78  &  0.67 & -  & 2.53 \\
$m$ FD, $m$ HD  & 18.15   & 1.37 & 2.59 &  2.54 \\
\hline
\hline
\multicolumn{5}{|c|}{$m=4$}\\
\hline
$2m$ FD  &  25.6 & -  & 1.6  &  -   \\
$2m$ HD   &  12.7  &  0.2 & -  & 1.39 \\
$m$ FD, $m$ HD & 18.37   & 0.5  & 1.38  & 1.32 \\
\hline
\hline
\end{tabular}
\caption{Goodputs with coexistence for multiple nodes, numbers taken from \cite{MelissaTVT}.}
\label{tbl:coexist_noeifs}
\end{table*}

We note from Figure 1(a) that the 802.11 system gives fewer time-slots for downlink than for uplink in a legacy half-duplex case. This motivates us to consider a system where the downlink and uplink throughput of half-duplex station is symmetric and also achieves fairness between half-duplex and full-duplex. Thus, we consider a system with $p_H=p_F=\frac{p_A}{n}=\frac{1}{2n+m}$ for $n>0$ and $p_F=\frac{1}{m}$, $p_H=p_A=0$ for $n=0$. With these settings, the uplink/downlink throughput from/to any of the  full- or half-duplex station is the same and equals $\frac{1}{2n+m}$. The system is thus fully-symmetrical in the throughput between the AP and any station. At the extreme scenarios of no full-duplex node, the access point gets the channel with probability $1/2$ and makes the uplink and downlink throughput symmetric. We note from Figure 1(b) that this symmetric rates come at a relatively moderate cost of total system throughput. The total system throughput reduces because AP grabs the channel more often thus favoring more half-duplex downlink throughput which further results in hurting the total system throughput. The adjustment of the probability to achieve fairness can be done without special changes to the stations, for example AP can access the channel with probability $p_A=\frac{n}{2n+m}$ and let the stations access the channel equally likely in the remaining time-slots. This could be achieved by adjusting the backoff mechanisms in DCA.



\section{Simulation Results}
In this section, we will describe our simulation results based on a mixed HD/FD system using a slightly-modified (for HD) random access 802.11 DCA scheme using backoff timers and various xIFS time spacings as well as RTS/CTS. The details of this MAC design for full-duplex nodes is described in \cite{MelissaTVT} where the co-existence of full-duplex and half-duplex nodes in a mixed HD/FD system was studied in Section VII.C and Section VII.D. We find that the results in Table \ref{tbl:coexist_noeifs} (based on Section VII.C of \cite{MelissaTVT}) are as predicted by the analytical results in the previous section since we have $p_A=p_F=p_H=\frac{1}{m+n+1}$. The total throughput in the simulation results, for a mix of $m$ FD and $m$ HD nodes, increases by a factor of $1.39$, $1.42$, and $1.45$ for $m=1, 2, 4$ respectively from the complete half-duplex system, as compared to $1.33$, $1.4$ and $1.44$ as predicted from the theory (which predicts a gain of $1+\frac{m}{2m+1}$). Thus, we see that the predicted mixed HD/FD throughput gains closely match the results from detailed simulations of a mixed HD/FD system based on 802.11 DCA for the special case of $p_A=p_F=p_H=\frac{1}{m+n+1}$. This paper derives the throughput for a mixture of full- and half- duplex nodes for any random access probability of nodes thus provides a thorough derivation of throughput observed in \cite{MelissaTVT}.

\section{Conclusion}

This paper considers a random-access time-slotted wireless network with a single access point and a mix of half- and full- duplex stations. We assume that all traffic is between the access point and stations, and also that the access point has a single queue and picks the packet at the head of the queue when it initiates a transmission, but can search the queue for a suitable packet when responding to an incoming full-duplex transmission. For the general case of different channel access probability for each class of nodes, we have provided generalized analytical formulations for the steady state saturated throughputs to/from each station. Full-duplex transmissions involve data transmitted simultaneously to and from the access point, and this influences the dynamics of the queue at the access point.
In an equal access system for all nodes as in 802.11 DCA, we find that the access point never initiates transmission to a full-duplex station when it wins the channel contention. We show that it is possible to achieve full fairness among all the four flows of half-duplex and full-duplex uplink and downlink by adjusting the probability of channel access at the access point.

\end{document}